\documentclass[prl,reprint,twocolumn,showpacs,preprintnumbers,amsmath,amssymb,nofootinbib,floatfix]{revtex4-1} 
\usepackage{graphicx}
\usepackage{mathrsfs}
\usepackage{hyperref}
\usepackage{slashed}
\usepackage{color}
\usepackage{booktabs}
\usepackage{ulem}
\usepackage{multirow}

\newcommand{\beq}{\begin{eqnarray}}
\newcommand{\eeq}{\end{eqnarray}}

\newcommand{\SC}[1]{{\color{blue}{#1}}}

\begin{document}

\title{Reviving the energy-dependent partonic structure of $f_0(980)$ via \\ two-pion distribution amplitudes}

\author{Shan Cheng}\email{Corresponding author: scheng@hnu.edu.cn}
\author{Ling-yun Dai}
\author{Jian-ming Shen}
\author{Shu-lei Zhang}\email{Corresponding author: zhangshulei@hnu.edu.cn}

\affiliation{School for Theoretical Physics, School of Physics and Electronics and Hunan Provincial Key Laboratory of High-Energy Scale Physics and Applications, Hunan University, 410082 Changsha, China}

\date{\today}

\begin{abstract}

We present a novel framework for analyzing four-body semileptonic weak decays, performing the first high-twist analysis of the $D_s \to \left[ \pi\pi \right]_{\rm S}$ form factors using light-cone distribution amplitudes for the isoscalar two-pion state ($2\pi$DAs). It is motivated by persistent tensions between phenomenological cascade analyses and QCD-based interpretations of the partonic structure of $f_0$ meson. Our study reveals that the twist-3 $2\pi$DAs incorporate an asymmetry in the partial-wave expansion, a feature absent in single $f_0$ distribution amplitudes, that drives a substantial cancellation between twist-2 and twist-3 contributions to the form factors. As a result, the predicted decay rate falls significantly below the experimental value. These findings indicate that the isoscalar two-pion system near the charm scale is not primarily a $q{\bar q}$ state, reviving the energy-dependent partonic structures of light scalar mesons by highlighting a fundamental limitation of the single-meson LCDA approach in describing such complex systems.
  
\end{abstract}


\maketitle

\textbf{\textit{Introduction.}}--The nature of light scalar mesons remains a long-standing puzzle in hadron physics \cite{ParticleDataGroup:2024cfk}. 
Experimental identification of these states is particularly challenging due to their characteristically large widths. 
From a quantum theory, the light scalar meson is a superposition of all possible Fock states
\beq \vert S \rangle = \psi_{q{\bar q}}\vert q {\bar q} \rangle + \psi_{q{\bar q}g} \vert q {\bar q} g \rangle 
+ \psi_{q {\bar q} q {\bar q}} \vert q {\bar q} q {\bar q} \rangle +  \cdots. \label{eq:Fockexpansion} \eeq 
The probability amplitudes $\psi_{n}$, representing different partonic configurations, depend on the partons' momentum distributions and their spin projections. 

Hadron spectroscopy offers clear evidence for such complex configurations \cite{Close:2002zu,Amsler:2004ps,Pelaez:2015qba}, fueling ongoing theoretical debates about the structure of both conventional resonances \cite{Bramon:1980ni,Morgan:1993td,Tornqvist:1995ay,Bicudo:1998mc,Umekawa:2004js}, which may mix with gluonball states \cite{RuizArriola:2010fj}, and exotic multi-particle states \cite{Oller:1997ti,Alford:2000mm,Achasov:2005hm}. In the case of exotic hadrons, the discussion centers on distinguishing between compact tetraquark states \cite{Jaffe:1976ig,Maiani:2004uc,tHooft:2008rus,Agaev:2017cfz,Briceno:2017qmb,Briceno:2016mjc} and loosely bound molecular states \cite{Weinstein:1982gc-1983gd-1990gu,Janssen:1994wn,Oller:1997ng,Locher:1997gr,Bernard:2010fp}.
While spectral analysis reveals these exotic configurations, the intrinsically nonperturbative nature of QCD prevents a direct extraction of the underlying parton dynamics.

\begin{figure}[t] \begin{center} \vspace{-2mm}
\includegraphics[width=0.35\textwidth]{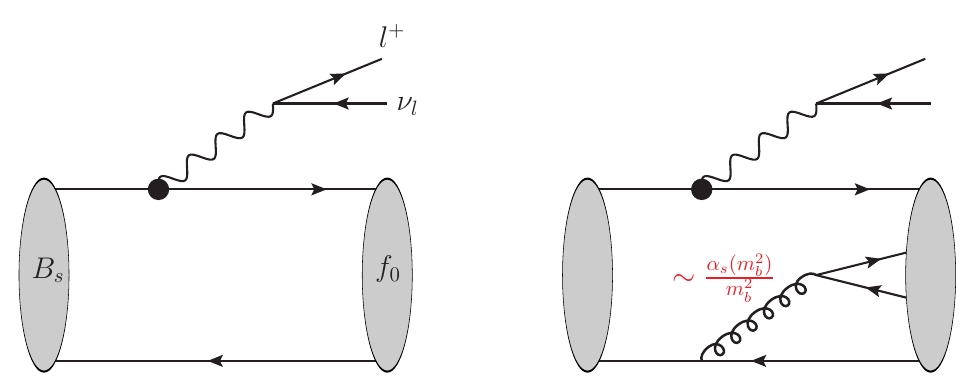} 
\end{center} \vspace{-4mm}
\caption{Illustrative diagrams of the color transparency mechanism in $B_s \to f_0 e^+ \nu_e$ decay.} \vspace{-4mm}
\label{fig:colortrans} \end{figure}

Semileptonic heavy meson decays are a powerful probe of scalar meson structure \cite{Wang:2009azc,Ke:2009ed,Sekihara:2015iha,Shi:2015kha,Cheng:2017fkw,Shi:2020rkz}, where the quark-gluon dynamics are encoded in light-cone distribution amplitudes (LCDAs) \cite{Lu:2006fr,Wu:2022qqx,Bijnens:2002mg,Ball:2006wn,Balitsky:1987bk,Braun:2003rp}. 
A key distinction arises between $B_s$ and $D_s$ decays due to the phenomenon of color transparency. In the decay $B_s \to f_0 \ell^+ \nu_\ell$, the energetic $f_0$ is dominated by its lowest Fock state \cite{Colangelo:2010bg,Cheng:2005nb,Cheng:2019tgh,Cheng:2023knr}, since the higher-state contributions are doubly suppressed by the large momentum transfer and the strong coupling constant, as depicted in Fig. \ref{fig:colortrans}. The $D_s \to f_0 \ell^+ \nu_\ell$ decay, however, produces a less relativistic quark pair. This makes the hadronization process more susceptible to contributions from higher Fock states, where the creation of additional gluons and $q\bar{q}$ pairs plays a significant role.

While the energy-dependent partonic structures is a QCD result, its practical application is often hindered by the complexity of multi-particle LCDAs. Consequently, in QCD-based approaches like Light-Cone Sum Rules (LCSRs), the  $f_0$ meson is typically approximated by its lowest Fock state. This simplification is generally reliable for $B_s$ decays, but becomes inadequate in $D_s$ decays where multi-particle Fock states contribute substantially. Surprisingly, recent phenomenological studies using the cascade decay framework \cite{Cheng:2023knr,Hu:2025ool}, under the $q{\bar q}$ ansatz, show good agreement with experimental data \cite{BESIII:2023wgr} for the four-body semileptonic decay $D_s \to \left[ \pi \pi \right]_{\rm S} \ell^+ \nu_\ell$ ($D_{l4}$). These analyses incorporate the $D_s \to f_0(980)$ form factors obtained from LCSRs with a Flatt\'e parameterization of the $f_0(980)$ resonance, which is subsequently coupled to the isoscalar $\pi\pi$ state. Given the scale dependence observed in the form factor calculations, specifically, the shift in dominance from the non-asymptotic twist-2 LCDA for $B_s \to f_0$ transition to the two-particle twist-3 LCDAs for $D_s \to f_0$ transition \cite{Cheng:2023knr}, the seemingly agreement in $D_{l4}$ decays is largely attributed to the cascade framework, where the Flatt\'e parameterization of the intermediate resonance severely disrupts the assessment of color transparency for partonic structures.

In this paper, we propose a novel framework to analyze $D_{l4}$ weak decays by introducing light-cone distribution amplitudes for the isoscalar two-pion state ($2\pi$DAs). The core concept is to calculate $D_s \to \left[\pi\pi\right]_{\rm S}$ form factor directly, bypassing the need for modeling the intermediate $f_0$ resonance. The apparent agreement within the cascade framework is demonstrated to be an artifact, arising from the symmetric two-particle twist-3 LCDAs of the single $f_0$ resonance. This is clarified by the fact that the corresponding LCDAs of the signal state $\left[\pi\pi\right]_{\rm S}$ are fundamentally asymmetric. Our findings indicate that the $q\bar{q}$ state contributes only modestly to $D_s \to [\pi\pi]_{\rm S} \ell \nu$ decay. Consequently, data from charm decays favor a multi-parton structure for the isoscalar two-pion system. 

\textbf{\textit{$2\pi$DAs of the isoscalar two-pion system}.}--The $\pi\pi$ invariant mass spectrum serves as the primary signal channel for observing the $f_0$ resonance, as the $KK$ channel is strongly suppressed by phase space. $2\pi$DAs provide one of the most general descriptions within QCD for energetic two-pion systems. 
They were initially developed at leading twist in the reaction $\gamma^\ast \gamma \to \pi\pi$ \cite{Diehl:1998dk,Polyakov:1998ze,Cheng:2019hpq}, and have recently been extended to twist-3 level in the hard exclusive processes \cite{Lorce:2022tiq,Cheng:2025hxe}. 
Twist is a quantum number that organizes the partonic configurations of the constituent quark and gluon fields.

For an isoscalar two-pion system, $2\pi$DAs of the lowest Fock state are defined by
\beq &&\langle \pi(k_1)\pi(k_2) \vert {\bar q}(0) \gamma^\mu q(x) \vert 0 \rangle = \int du e^{i {\bar u} k \cdot x} 
k^\mu \Phi_\parallel, \nonumber\\
&&\langle \pi(k_1)\pi(k_2) \vert {\bar q}(0) q(x) \vert 0 \rangle = \int du e^{i {\bar u} k \cdot x} 
\frac{i k^2 (k \cdot x) }{2 f_{2\pi}^\perp} \Phi_\parallel^{(s)}, \nonumber\\
&&\langle \pi(k_1)\pi(k_2) \vert {\bar q}(0) \sigma^{\mu\nu} q(x) \vert 0 \rangle =  - \frac{i}{f_{2\pi}^\perp} \int du e^{i {\bar u} k \cdot x} \nonumber\\
&& \hspace{1.4cm} \Big[ \frac{k_\mu {\bar k}_\nu - k_\nu {\bar k}_\mu }{2 \zeta -1} \Phi_\perp - k^2 \frac{k_\mu x_\nu - k_\nu x_\mu}{k \cdot x} \Phi_\parallel^{(t)} \Big].
\label{eq:2pidas-definition} \eeq
Here $\Phi_\parallel, \Phi_\perp$ and $\Phi_\parallel^{t}, \Phi_\parallel^{s}$ are leading and subleading twsit $2\pi$DAs, respectively. $k = k_1 + k_2$ (${\bar k} = k_1 - k_2$) denotes the sum (difference) of two pions' momenta, $f_{2\pi}^\perp$ is the decay constant defined via the local matrix element of the tensor current in the soft-meson limit. 
$2\pi$DAs are dimensionless functions of three variables: 
the longitudinal momentum fraction $u$ (${\bar u}$) carried by the antiquark (quark) field relative to $k$, the single pion momentum fraction $\zeta = k_1^+/k^+$ along the positive direction, and the invariant mass squared $k^2$. They can be expressed formally as a double expansion in Gegenbauer $C_n^{3/2}(2 u-1)$ and Legendre polynomials $C_l^{1/2}(2\zeta-1)$, by combining QCD's conformal symmetry with partial wave analysis.
\beq &&\Phi_{\parallel,\perp} = 6u{\bar u} \, C_1^{3/2} 
\Big[ B_{10}^{\parallel,\perp}(k^2, \mu) C_0^{1/2} + B_{12}^{\parallel,\perp}(k^2, \mu) C_2^{1/2} \Big], \nonumber\\
&&\Phi^{(s)}_{\parallel} = 6u{\bar u} \, C_1^{3/2} 
\Big[ B_{10}^\parallel(k^2, \mu) C_0^{1/2} + B_{12}^\parallel(k^2, \mu) C_2^{1/2} \Big], \nonumber\\
&&\Phi^{(t)}_{\parallel} = C_1^{3/2} \Big[ B_{10}^\parallel(k^2, \mu) C_0^{1/2} + B_{12}^\parallel(k^2, \mu) C_2^{1/2} \Big]. \label{eq:2pidas-t3} \eeq
The relation $\Phi(u, \zeta, k^2) = - \Phi(1-u, \zeta, k^2)$, constrained by the $C$-parity, is maintained by the Gegenbauer polynomials with odd number $n$. 

The expansion coefficients $B_{nl}$ weights the universal DAs at a given conformal power $n$ and partial wave $l$. 
\beq B^{\parallel}_{nl}(\mu, k^{2}) 
&=& B^{\parallel}_{nl}(0) \left[ \frac{\alpha_s(\mu)}{\alpha_s(\mu_0)}\right]^{\frac{\gamma_n - \gamma_0}{2\beta_0}} {\rm Exp} \Big[ k^{2} \frac{d \ln B^{\parallel}_{nl}(0)}{dk^{2}} \nonumber\\
&~& + \frac{k^{4}}{\pi} \int_{4m_\pi^2}^\infty ds \frac{\delta_l(s)}{s^2 (s-k^{2}-i0)} \Big]. \label{eq:Bnl} \eeq 
From one hand, these coefficients incorporate the scale dependence through renormalization group evolution  \cite{Ball:1996tb,Ball:1998sk,Ball:1998ff}, 
mirroring the well-established behavior of single-meson LCDAs governed by one-loop anomalous dimensions $\gamma_n$ \cite{Chernyak:1977as,Lepage:1979zb,Efremov:1979qk}. 
From the other hand, they evolve with $k^2$, thereby capturing contributions from both intermediate resonant states and the nonresonant QCD background. For the resonant regions, the $k^2$ dependence can be obtained via a subtracted dispersion relation, guided by Watson's theorem \cite{Omnes:1958hv}, and constrained by data on phase shifts \SC{$\delta_l$} from $\pi N$ scattering \cite{GAMS:1994jtk,Achasov:1998pu} and $\pi\pi$ scattering \cite{Garcia-Martin:2011iqs,Dai:2014zta}. 
The initial values of the subtracted coefficients in the threshold region are derived from low-energy theorems based on the instanton vacuum \cite{Diakonov:1985eg,Lehmann-Dronke:1999vvq}, as provided in the Supplement Material \cite{Supplement} and the references therein. For the leading-twist $2\pi$DAs, this approach yields the coefficients at the first power: $B_{10}^\parallel(0) = - B_{12}^\parallel(0) = -0.556$, $d \ln B^{\parallel}_{nl}(0)/dk^{2} = 0.413 \, {\rm GeV}^{-2}$. 

\begin{figure*}[t] \begin{center} \vspace{-2mm}
\includegraphics[width=0.30\textwidth]{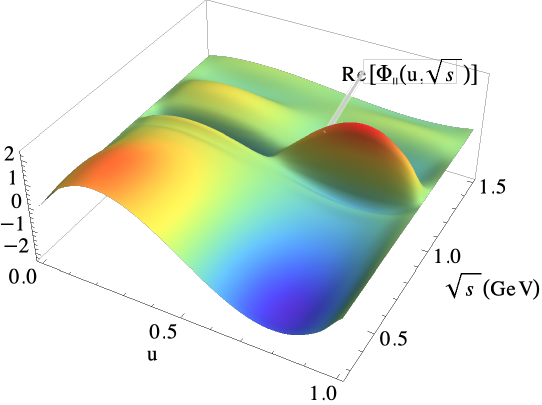}  \hspace{2mm}
\includegraphics[width=0.03\textwidth]{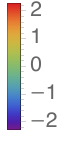} \hspace{6mm}
\includegraphics[width=0.30\textwidth]{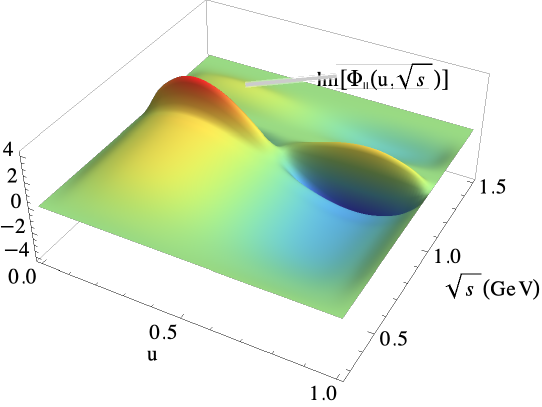} \hspace{2mm}
\includegraphics[width=0.03\textwidth]{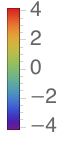} \vspace{6mm}
\includegraphics[width=0.30\textwidth]{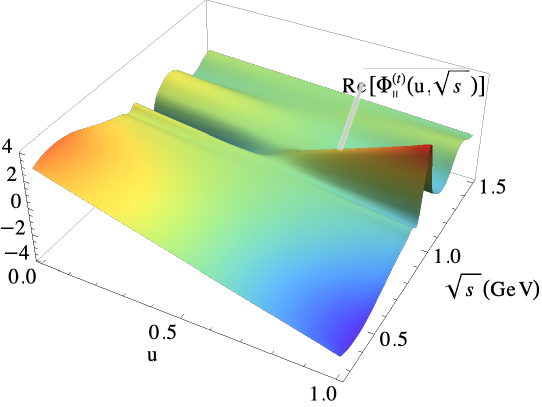}  \hspace{2mm}
\includegraphics[width=0.03\textwidth]{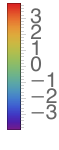} \hspace{6mm}
\includegraphics[width=0.30\textwidth]{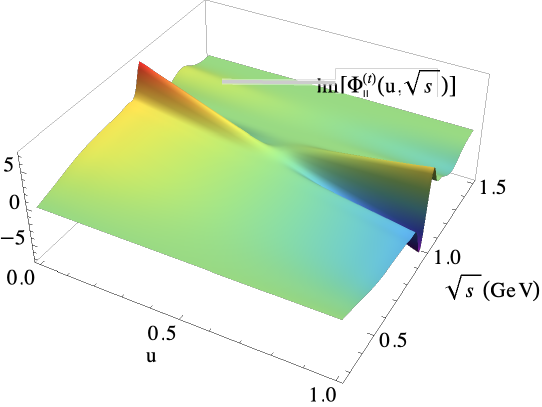} \hspace{2mm}
\includegraphics[width=0.04\textwidth]{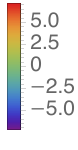} 
\end{center} \vspace{-8mm}
\caption{Twist-2 (up) and twist-3 (low) isoscalar $2\pi$DAs at the scale $\mu=1.3 \, {\rm GeV}$, by taking momentum fraction $\zeta=0.5$.}
\label{fig:Phipara}
\end{figure*} 

In Fig. \ref{fig:Phipara}, we present two-dimension plots of the twist-2 ($\Phi_\parallel$) and twist-3 ($\Phi_\parallel^{(t)}$) $2\pi$DAs as functions of the parton momentum fraction $u$ and the invariant mass $\sqrt{s}=k$. The pion momentum fraction is fixed at $\zeta=0.5$ and the renormalization scale is typically set at $\mu=1.3$ GeV for charm-meson decays. Three key pieces of information can be gleaned from the figure. First, a distinct structural feature, either a peak or a dip, emerges in the $f_0(980)$ region, whereas no comparable signal appears in the $\sigma(500)$ region. This is consistent with the flat phase shift behavior observed in $\pi\pi$ scattering \cite{Garcia-Martin:2011iqs,Dai:2014zta}. Second, the twist‑2 and twist‑3 LCDAs display asymmetry in both their real and imaginary parts with respect to the midpoint of the parton momentum fraction. This contrasts sharply with the LCDAs of a single single $f_0$ meson, for which the twist‑3 DAs are symmetric about $u=0.5$ \cite{Cheng:2023knr,Hu:2025ool}. Finally, twist-3 LCDA $\Phi_\parallel^{(t)}$ displays end-point enhancements, in contrast to the vanishing of the leading-twist LCDA at the boundaries. As we will see, the last two features, especially the second one, predominantly explain the non-$q\bar{q}$ component of the isoscalar two-pion system  in $D_{l4}$ decays.

\textbf{\textit{$D_{s} \to \left[ \pi\pi \right]_{\rm S}$ form factors and the $D_{l4}$ decays.}}--The transition from a $D_{s}$ meson to a two-pion system can be described by the matrix element expressed through four orthogonal form factors $F_{i=\parallel, \perp, 0, t}$ \cite{Faller:2013dwa,Kang:2013jaa,Meissner:2013hya,Boer:2016iez}, 
\beq &~&\langle \pi^+(k_1) \pi^-(k_2) \vert {\bar s} \gamma^\mu \left( 1 - \gamma_5 \right) b \vert D_{s}^+(p) \rangle \nonumber\\
&=& i F_\perp \frac{{\bar q}_\perp^\mu}{\sqrt{k^2}} + F_t \frac{q^\mu}{\sqrt{q^2}} + F_0 \frac{2 \sqrt{q^2}  k_0^\mu}{\sqrt{\lambda}} + F_\parallel \frac{{\bar k}^\mu_{\parallel}}{\sqrt{k^2}}. \label{eq:B2pipi-ff} \eeq
The momentum basis are ${\bar q}_\perp^\mu = 2 \varepsilon^{\mu\nu\rho\sigma}q_\nu k_\rho {\bar k}_\sigma/\sqrt{\lambda}$, $q^\mu = p^\mu - k^\mu$, $k_0^\mu = k^\mu - k\cdot q/q^2 q^\mu$, ${\bar k}^\mu_{\parallel} = {\bar k}^\mu - 4 (k \cdot q) ({\bar k} \cdot q)\lambda k^\mu + 4 k^2 ({\bar k} \cdot q)/\lambda q^\mu$. Form factors $F_{i=\parallel, \perp, 0, t}$ are functions of three kinematical variables, saying the momentum transfer squared $q^2$, the two-pion invariant mass squared $k^2$, and the scalar produce $q \cdot {\bar k} = \beta_\pi\cos \theta_\pi \sqrt{\lambda}/2$ that specifies the polar angle $\theta_\pi$ of the $\pi^+$ meson in the two-pion rest frame. In addition, $\lambda = m_B^4 - q^4 - k^4 - 2 ( m_B^2 q^2 + m_B^2 k^2 + q^2 k^2 )$ is the K\"all\'en function, $\beta_\pi = ( 1 - 4 m_\pi^2/s )^{1/2}$ is the phase space factor. 

The LCSRs' derivation of $D_{s} \to \pi\pi$ form factors begins with a twofold analysis of the two-point correlation function, defined as the vacuum-to-two-pion matrix element of the time-ordered product of the bilocal currents $j_\mu = {\bar s} \gamma_\mu (1-\gamma_5) c$ and $j_{D_{s}} = i m_c {\bar c} \gamma_5 s$. In the negative half-plane of momentum transfers, it can be computed directly via the operator product expansion, where it is expressed as a convolution of perturbative scattering amplitudes with LCDAs, organized by twist. When $q^2$ crosses into positive half-plane, the dynamics shift to long-distance quark–gluon interactions, and hadronic states begin to form. In this regime, the correlator is instead represented as a sum over contributions from all possible intermediate hadronic states. The final result is obtained by applying quark-hadron duality where the ground state is separated out from the interpolating current. The longitudinal form factors that contribute to $D_{(s)} \to \left[ \pi\pi \right]_{\rm S}$ transition ultimately read as \cite{Cheng:2025hxe} 
\begin{widetext} \vspace{-2mm} \beq
&&\sqrt{q^2} F_0^{(l'=0,2,\cdots)}(q^2, k^2) = \frac{\sqrt{\lambda}}{m_{D_{s}}^2-q^2-k^2} \sqrt{q^2} F_t^{(l')}(q^2, k^2) \nonumber\\
&& \hspace{1cm} - \frac{i m_c k^2 q^2}{2m_{D_{s}}^2 f_{D_{s}} f_{2\pi}^\perp} \frac{\sqrt{\lambda}}{m_{D_{s}}^2-q^2-k^2} 
\int_{u_0}^1 du \left( \frac{1}{u^2} - \frac{m_c^2-q^2+u^2k^2}{u^3M^2} \right) \Phi^{\parallel, ({\rm st})}(u) \left[ B_{10}^\parallel I^t_{0l'} + B_{12}^\parallel I^t_{2l'} \right] e^{\frac{m_{D_{s}}^2-s(u,q^2)}{M^2}} \nonumber\\
&&\hspace{1cm}+ \frac{i m_c k^2 q^2}{2m_{D_{s}}^2 f_{D_{s}} f_{2\pi}^\perp} \frac{\sqrt{\lambda}}{m_{D_{s}}^2-q^2-k^2} \frac{m_c^2-q^2+u_0^2k^2}{u_0 (s_0 - q^2)} \Phi^{\parallel, ({\rm st})}(u_0) \left[ B_{10}^\parallel I^t_{0l'} + B_{12}^\parallel I^t_{2l'} \right] e^{\frac{m_{D_{s}}^2-s_0}{M^2}}, \label{eq:F0-2} \eeq \vspace{-4mm}\end{widetext}
where the auxiliary distribution amplitude $\Phi^{\parallel, ({\rm st})}(u) = - 9 u {\bar u} (2u-1) + 6 u^2 - 6 u$ incorporates the sum of the earilier known twist-2 contribution \cite{Hambrock:2015aor, Cheng:2017sfk, Cheng:2017smj} and the recently calculated twist-3 contribution \cite{Cheng:2025hxe}. The threshold momentum fraction $u_0$ is the solution of $s_0 = {\bar u} k^2 + (m_c^2 + {\bar u} q^2)/u$. Moreover, the integration over the polar angle $\theta_\pi$ is given by $I_{ll'}^t= \int_{-1}^1 d(\cos\theta_\pi) P_{l'}^{(0)}(\cos\theta_\pi) P_{l}^{(0)}(\beta_\pi \cos\theta_\pi)$.

We adopt the PDG values of $m_{D_s}= 1.968$ GeV and $f_{D_s}= 0.250$ GeV \cite{ParticleDataGroup:2024cfk}. For the LCSRs parameters, we choose $M^2 = 5.0 \pm 0.5$ GeV$^2$ and $s_0 = 6.0 \pm 0.5$ GeV$^2$, in accordance with the earlier study of $D_s \to f_0$ form factors \cite{Cheng:2023knr}, to ensures optimal stability and convergence of the QCD predictions. Charm quark mass and nonpertuabtive parameters in $2\pi$DAs are taken from Ref. \cite{Cheng:2025hxe}. The phase shifts $\delta_l(k^2)$ incorporated into the expansion coefficients in Eq. (\ref{eq:Bnl}) is taken from the pion-nucleon reaction $\pi^- p \to \pi^0\pi^0 n$ \cite{GAMS:1994jtk,Achasov:1998pu}. This choice of reaction is motivated by the similarity of its peak structure to that observed for the $f_0$ produced in $D_s$ decays. 

Table \ref{tab:twist-clarification} presents the contributions from different twists to the $S$- and $D$-wave $D_s \to [\pi\pi]_{\rm S}$ form factors at full recoil ($q^2 = 0$), obtained after integrating over the $\pi\pi$ invariant mass $k^2$. For comparison, the corresponding $D_s \to f_0$ form factor is also included. While the $D_s \to f_0$ form factor \cite{Cheng:2023knr} is real, the $D_s \to [\pi\pi]_{\rm S}$ form factors are complex. Furthermore, in contrast to the LCSR result for $D_s \to f_0$ form factors where twist-2 and twist-3 contributions both provide enhancement, the direct calculation for the $D_s \to [\pi\pi]_{\rm S}$ transition reveals a non-trivial and instructive interplay between them. The origin of this effect lies in the asymmetry of the twist-3 $2\pi$DAs exhibited earlier. This contrasts with the symmetric twist-3 LCDAs of $f_0$, where QCD sum rules dictate that the asymmetric terms vanish, leaving only the symmetric terms associated to the even Gegenbauer polynomial \cite{Cheng:2005nb,Cheng:2019tgh,Cheng:2023knr}.

\begin{table}[t] \vspace{-4mm}
\caption{The central values of twist-2 and twist-3 contributions to $D_s \to \left[ \pi\pi \right]_{\rm S}$ form factors at the full recoiled point $q^2 = 0$, in comparing to the $D_s \to f_0$ form factor.} \vspace{2mm}
\renewcommand{\arraystretch}{1.4}   
\begin{tabular}{c | c | c | c } 
\hline\hline 
\textbf{FFs} \quad & \quad $\sqrt{q^2} F_0^{(l=0)}(0)$ \quad & \quad $\sqrt{q^2} F_0^{(l=2)}(0)$ \quad & \quad $f_+(0)$ \quad \\ \hline
\textbf{Twist-2} \quad & \quad $0.20 - i 0.24$ \quad & \quad $0.27 + i 0.21$ \quad & \quad $0.20$ 
\quad \nonumber\\ \hline
\textbf{Twist-3} \quad & \quad $-0.41 + i 0.51$ \quad & \quad $-0.55 - i 0.41$ \quad & \quad $0.41$
\quad \nonumber\\ \hline
\textbf{Total} \quad & \quad $-0.21 + i 0.27$ \quad & \quad $-0.28 - i 0.20$ \quad & \quad $0.61$ \quad \nonumber\\
\hline\hline  
\end{tabular} \label{tab:twist-clarification}
\end{table} 

\begin{figure}[b]
\begin{center}  
\includegraphics[width=0.20\textwidth]{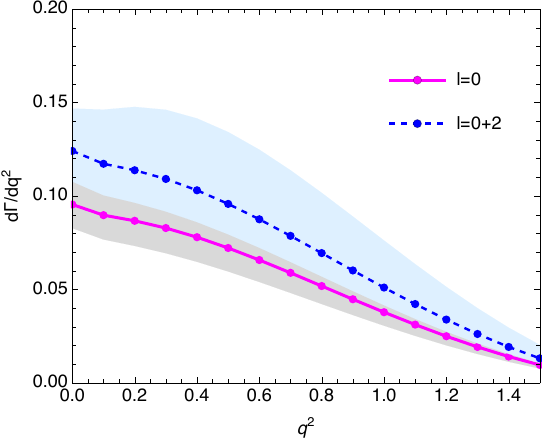}  \hspace{4mm}
\includegraphics[width=0.20\textwidth]{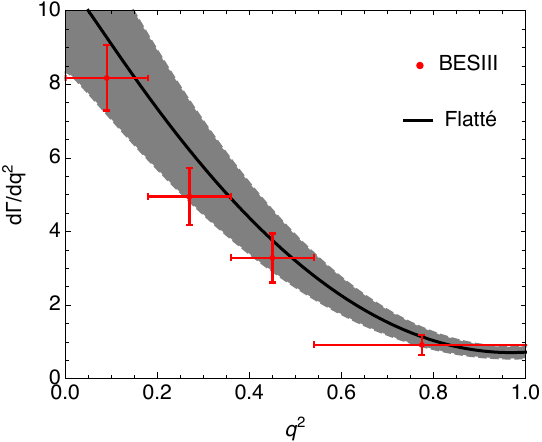} 
\end{center} \vspace{-4mm}
\caption{The differential width $d\Gamma/dq^2$ of $D_{(s)} \to \left[ \pi\pi \right]_{\rm S} e^+ \nu_e$ (left)
and $D_s \to \left[ f_0 \to \pi\pi \right] e^+\nu_e$ decays (right).}
\label{fig:dGammadq2}
\end{figure}

The differential width of $D_s \to \left[ \pi\pi \right]_{\rm S} e^+ \nu_e$ decay on the momentum transfers is quoted as \cite{Cheng:2025hxe} 
\beq \frac{d\Gamma}{dq^2} = \int_{4m_\pi^2}^{k^2_{\rm max}} dk^2 \frac{ G_F^2 \vert V_{cs} \vert^2 \beta_\pi \sqrt{\lambda} q^2}{3 \left( 4\pi \right)^5 m_{D_s}^3} \sum_{l} \vert F_0^{(l)}(q^2,k^2) \vert^2 . 
\label{eq:Bl4-2D} \eeq 
Here the integral limit $k^2_{\rm max}$ is a $q^2$-function related to the kinematics. We employ the simplified $z$-series expansion  parameterization \cite{Bourrely:2008za} to extend the LCSRs prediction in the large-recoil region, approximately $0 \lesssim q^2 \lesssim 0.2$ GeV$^2$, to across the entire physical range. Fig. \ref{fig:dGammadq2} depicts the result. In the left panel, the $S$- and $D$-wave contributions to the $D_{l4}$ decay are displayed with iteration. The right panel compares BESIII data \cite{BESIII:2023wgr} to the prediction derived under the cascade framework \cite{Cheng:2023knr} where the decay $D_s \to f_0 e^+ \nu_e$ is followed by $f_0 \to \pi\pi$, employing a Flatt\'e parameterization for the resonance. Tab. \ref{tab:Gamma} provides the decay widths obtained by integrating $q^2$ within these two frameworks under the $q{\bar q}$ ansatz.

While the cascade decay analysis can reproduce the BESIII data, it critically depends on two artificial assumptions: a marked sensitivity to the parameterization of the resonance, and the theoretically unsound use of a single-meson LCDAs for the $f_0$. The observed agreement therefore appears to arise from a fine-tuned, and presumably coincidental, delicate interplay between these methodological choices. In contrast, the result of a direct $D_{l4}$ analysis using $D_s \to [\pi\pi]_{\rm S}$ form factors is roughly two orders of magnitude smaller than the BESIII data. Such a significant discrepancy reveals that, unlike in analogous $B$-meson decays \cite{Cheng:2025hxe}, the isoscalar two-pion system at the charm scale is not primarily a $q\bar{q}$ state. This result provides strong support for the partonic configuration interpretation governed by color transparency mechanism. 

\begin{table}[t] \vspace{-4mm}
\caption{Decay widths (in unit of $10^{-4}$) obtained from direct $D_{l4}$ framework and cascade decay framework.} \vspace{2mm} 
\renewcommand{\arraystretch}{1.4} 
\begin{tabular}{ c | c | c } 
\hline\hline
$D_s \to \left[ \pi\pi \right]_{\rm S} e^+ \nu_e$ \quad & \quad $D_s \to \left[ f_0 \to \pi\pi \right] e^+ \nu_e$  \cite{Cheng:2023knr} \quad & 
\quad Data \cite{BESIII:2023wgr} \\ \hline
$0.81^{+0.34}_{-0.14}$ & $18.8^{+4.5}_{-3.8} $ & $17.2 \pm 1.6 $ \nonumber\\ 
\hline\hline
\end{tabular} \label{tab:Gamma}
\end{table} 

\textbf{\textit{Conclusions.}}--Introducing $2\pi$DAs to describe the partonic structure of the isoscalar two-pion state, the visible signal channel of the $f_0$ meson, we propose a novel framework for analyzing $D_{l4}$ decays through the direct calculation of $D_s \to \left[ \pi\pi \right]_{\rm S}$ form factors. We observe a significant cancellation between the contributions from twist-2 and twist-3 $2\pi$DAs, leading to a predicted decay width significantly smaller than the BESIII measurement. This outcome aligns with QCD-based expectations from the color transparency mechanism, supporting the view that the $f_0$ meson at the charm mass scale is not dominated by the lowest $q{\bar q}$ Fock state. These findings emphasize a fundamental limitation of the traditional single-particle LCDAs framework when it comes to characterizing the structure of light scalar meson. In such cases, the asymptotic term vanishes, and twist-3 distribution amplitudes become dominant. These higher-twist amplitudes encode an asymmetry in the partial-wave expansion that the single-particle picture fails to capture. 
Ultimately, our results reveal the rich and scale-dependent structure of $f_0$ in exclusive processes, underscoring the necessity of a multi-particle, channel-sensitive approach to describe resonances in QCD.
  
{\it Acknowledgments:} We are grateful to Xian-wei Kang, Hai-bo Li and Wei Wang for insightful discussions. Special thanks go to Hai-yang Cheng for reviewing an earlier draft of the manuscript and for his constructive comments. This work is supported by the National Key R$\&$D Program of China under Contracts No. 2023YFA1606000 and the National Science Foundation of China (NSFC) under Grant No 12575098.

\pagebreak
\widetext
\begin{center} 
\vspace{6mm}
\textbf{\large Supplemental Materials to "Reviving the energy-dependent partonic structure of $f_0(980)$ via two-pion distribution amplitudes"} 
\end{center}
\setcounter{equation}{0}
\setcounter{figure}{0}
\setcounter{table}{0}
\setcounter{page}{1}
\makeatletter
\renewcommand{\theequation}{S\arabic{equation}}
\renewcommand{\thefigure}{S\arabic{figure}}
\renewcommand{\bibnumfmt}[1]{[S#1]}

\begin{appendix}

\vspace{-4mm}
\section{Determination of the nonperturbative parameters in isoscalar $2\pi$DAs}\label{app-lcdas}

The values of the subtracted coefficients in the threshold region are initially determined by low-energy theorems derived from the instanton vacuum model \cite{Diakonov:1985eg,Lehmann-Dronke:1999vvq}. Instantons represent nonperturbative fluctuations of the gluon field in QCD after appropriate smearing. This vacuum is modeled as a grand canonical ensemble, with its partition function expressed in terms of collective coordinates. In Euclidean formalism, it is given by \cite{Pobylitsa:1989uq}: 
\beq Z = \int {\cal D} B_\mu \, e^{- \frac{1}{2g^2(M)} \int d^4x \, {\rm Tr} F_{\mu\nu}^2(A)} \int d \psi \, {\cal D} \psi^\dag \, e^{\int d^4x \, \psi^\dag \left( i \nabla A + i m \right) \psi}. \label{eq:partitionfunction}\eeq
Here, the gluon field is decomposed as $A_\mu(x) = {\bar A}(x,\gamma_{\rm I}) + B(x)$, where $B$ is the quantum field and ${\bar A}$ is the classical background field. 
This background is a superposition of istantons (${\rm I}$) and anti-instantons ($\overline{\rm I}$), with the particle numbers $N_+$ and $N_-$, respectively. The symbol $\gamma_{\rm I}$ denotes the collective coordinates, the center coordinate $z_{{\rm I} \mu}$, the size $\rho_{\rm I}$ and the $SU_{N_c}$ orientation matrix $U_{\rm I}$, of the ${\rm I}$th pseudoparticle. The total number of pseudoparticles is $N = N_+ + N_-$. Additionally, $\psi$ represents the light quark field. 

Within this framework, the mean values of the collective coordinates are obtained by maximizing the partition function for a fixed $N$. The results indicate that the instanton medium is relatively dilute, characterized by an average instanton size $\bar{\rho} \approx \left( 600 \, {\rm MeV} \right)^{-1}$ and an average separation $\bar{R} \approx \left( 200 \, {\rm MeV} \right)^{-1}$. Shuryak argued that such a dilute medium can explain both the gluon and quark condensates, along with numerous other observables \cite{Diakonov:1985eg}. A key element of this derivation is the introduction of a dimensionless parameter that incorporates the inverse coupling argument, a quantity not uniquely fixed from first principles. As noted in Ref. \cite{Diakonov:1995qy} (detailed in Tab. I thereby), the uncertainties in the extracted values of $\bar{\rho}$ and $\bar{R}$ are each approximately $10\%$. Their ratio $\bar{\rho}/\bar{R} \sim 1/3$, however, proves notably stable against variations in this parameter. In our analysis, we adopt the well‑established values from Refs. \cite{Shuryak:1982dp,Petrov:1998kg}, allowing the associated uncertainties to be safely neglected in the construction of the $2\pi$DAs.

The effective low-energy theory of pion interactions with massive constituent quarks, derived from the instanton model of the QCD vacuum in the large $N_c$ limit \cite{Diakonov:1985eg}, is formulated with a coupling governed by the action \cite{Petrov:1998kg}
\beq S_{\rm eff} = \int d^4x {\bar \psi}(x) \left[ i \hat{\partial} - M F(\partial^2) e^{\frac{i \gamma_5 \tau^a \pi^a(x)}{f_\pi}} F(\partial^2) \right] \psi(x). 
\label{eq:quark-pion-coupling}\eeq
Here, $M$ represents the momentum-dependent dynamical quark mass generated in the spontaneous breaking of chiral symmetry, which serves as an effective UV regulator for loop integrals by cutting them off at momenta on the order of the inverse average instanton size ${\bar \rho}^{-1} \approx 600$ MeV. Their ratio is proportional to the packing fraction of the instanton mesium $\left( M {\bar \rho} \right)^2 \sim \left( {\bar \rho}/{\bar R} \right)^4$. At zero momentum, it was determined to be $M(0) = 345$ MeV \cite{Diakonov:1985eg}, a scale that is parametrically small compared to ${\bar \rho}^{-1}$. The pion decay constant is taken as $f_\pi = 130$ MeV \cite{ParticleDataGroup:2024cfk}. 

\begin{figure}[h] \begin{center} 
\includegraphics[width=0.35\textwidth]{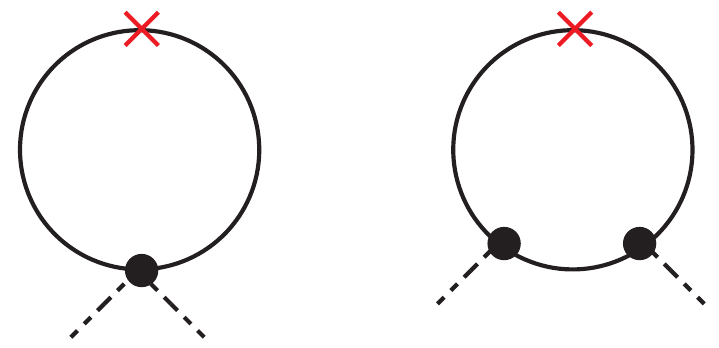} 
\end{center} \vspace{-4mm}
\caption{The typical diagrams contributed to $2\pi$DAs in the low-energy effective theory.} 
\label{fig:2piDAs} \end{figure} 

The general framework for calculating parton distribution amplitudes (DAs) within the low-energy effective theory was initially developed in Refs. \cite{Diakonov:1996sr,Diakonov:1997vc,Petrov:1998kg} and later extended to two-pion DAs in Refs. \cite{Polyakov:1998td,Polyakov:1999gs}.
Fig. \ref{fig:2piDAs} illustrates two representative diagrams contributing to the $2\pi$DAs, obtained by expanding the exponential in powers of the pion field. The dashed lines correspond to pion fields, the solid lines represent the massive quark propagator $1/[ i \hat{\partial} - M F^2(\partial^2) ]$, and the filled circles denote the quark-pion vertex derived from the effective action in Eq. (\ref{eq:quark-pion-coupling}), which introduces a form factor 
$F(\partial^2)$ for each quark line. It should be noted that the left diagram contributes only to the isoscalar distribution, whereas the right diagram contributes to both the isoscalar and isovector two-pion DAs.
The corresponding Feynman integrals can be evaluated directly by introducing light-cone coordinates. In this approach, the integral over transverse momenta exhibits a logarithmic divergence that is either regulated by the form factor
$F(\partial^2)$ or absorbed into the pion decay constant.

For the isoscalar $2\pi$DAs considered here, this approach yields the coefficients at leading power \cite{Polyakov:1998ze}: $B_{10}^\parallel(0) = - B_{12}^\parallel(0) = -0.556$, $d \ln B^{\parallel}_{nl}(0)/dk^{2} = 0.375 \, {\rm GeV}^{-2}$. 
The chirally odd $2\pi$DAs involve the decay constant, 
defined through the local matrix element $\langle \pi^+(k_1)\pi^-(k_2) \vert {\bar q} \sigma^{\mu\nu} q \vert 0 \rangle \stackrel{k^2 \to 0}{\longrightarrow} -i \left( k_\mu {\bar k}_\nu - k_\nu {\bar k}_\mu \right)/(2 f_{2\pi}^\perp)$, giving $f_{2\pi}^\perp(1 \, {\rm GeV}) \approx 0.567$ MeV. 
These parameters are flavor-universal under the $SU(3)$ limit since they are defined in terms of constituent quark fields via Eq. (\ref{eq:quark-pion-coupling}). Flavor dependence instead arises through the matrix elements of the quark operators in Eq. (2) of the main text. This dependence manifests in the $k^2$-evolution across the resonant region and is directly observable in $\pi\pi$ scattering and $\pi\pi-KK$ rescattering data. Gluonic $2\pi$DAs, defined by replacing quark fields with gluon fields in Eqs. (2), are parametrically suppressed at a low normalization scale $\mu \sim 1/\bar{\rho}$ by the instanton packing fraction. As a result, they vanish within the asymptotic approximation \cite{Diakonov:1995qy,Balla:1997hf}, consistent with the behavior previously discussed for three- and four-particle configurations.

Furthermore, $C$-parity indicates that isoscalar $2\pi$DAs possess the antisymmetry property: $\Phi^{I=0}(u, \zeta, k^2) = - \Phi^{I=0}({\bar u}, \zeta, k^2)$. At leading order in chiral perturbation theory, the scalar form factor in the low-energy region $\sqrt{k^2} \leqslant 0.1,\text{GeV}$ is given by $\Gamma_\pi(s) = m^2 + \mathcal{O}(p^4)$ \cite{Donoghue:1990xh}. Here, $m$ denotes the chiral symmetry-breaking term proportional to quark masses, and the expansion parameter satisfies $p/\Lambda \sim p/(4 \pi f_\pi) \leqslant 0.3$. 
For applications in heavy-flavor decays, the relevant kinematic domain lies in the physical region $\sqrt{k^2} > 2m_\pi$. Therefore, the definitions of $2\pi$DAs in Eqs. (2) are formulated in the chiral limit ($m_q \to 0$). Consequently, the scalar form factor obeys the normalization condition: $ \langle \pi(k_1)\pi(k_2) \vert m_q {\bar q}(0) q(x) \vert 0 \rangle \stackrel{x \to 0}{\longrightarrow} \Gamma_\pi(k^2) \stackrel{m_q \to 0}{\longrightarrow} 0$.


\end{appendix}

\end{document}